\documentclass[aps,pra,reprint,superscriptaddress,showpacs]{revtex4-1}
\usepackage[utf8]{inputenc}
\usepackage{graphicx}
\usepackage{amssymb,amsmath,amsthm,amsfonts}
\usepackage{hyperref}
\usepackage[capitalise,compress]{cleveref}
\usepackage{bm}
\usepackage{subfloat}
\usepackage{float}
\usepackage{color}
\usepackage[caption=false,justification=justified]{subfig}
\usepackage{xcolor}
\newcommand{\ket}[1]{\vert #1\rangle}

\newcommand{\comment}[1]{\textit{\color{gray} }}

\begin{document}

\title{Efficient, stabilized two-qubit gates on a trapped-ion quantum computer}
\author{Reinhold Bl\"umel}
\affiliation{Wesleyan University, Middletown, CT 06459, USA}
\affiliation{IonQ, College Park, MD 20740, USA}
\author{Nikodem Grzesiak}
\affiliation{IonQ, College Park, MD 20740, USA}
\author{Nhung H. Nguyen}
\affiliation{Joint Quantum Institute, University of Maryland, College Park, MD 20742, USA}
\author{Alaina M. Green}
\affiliation{Joint Quantum Institute, University of Maryland, College Park, MD 20742, USA}
\author{Ming Li}
\affiliation{IonQ, College Park, MD 20740, USA}
\author{Andrii Maksymov}
\affiliation{IonQ, College Park, MD 20740, USA}
\author{Norbert M. Linke}
\affiliation{Joint Quantum Institute, University of Maryland, College Park, MD 20742, USA}
\author{Yunseong Nam}
\affiliation{IonQ, College Park, MD 20740, USA}
\date{\today}

\begin{abstract}

Quantum computing is currently limited by the cost of two-qubit entangling operations. In order to scale up quantum processors and achieve a quantum advantage, it is crucial to economize on the power requirement of two-qubit gates, make them robust to drift in experimental parameters, and shorten the gate times. In this paper, we present two methods, one exact and one approximate, to construct optimal pulses for entangling gates on a pair of ions within a trapped ion chain, one of the leading quantum  computing architectures. Our methods are direct, non-iterative, and linear, and can construct gate-steering pulses requiring less power than the standard method by more than an order of magnitude in some parameter regimes. The power savings may generally be traded for reduced gate time and greater qubit connectivity. Additionally, our methods provide increased robustness to mode drift. We illustrate these trade-offs on a trapped-ion quantum computer.
\end{abstract}


\maketitle

\noindent 

{\it Introduction}.---For the growing number of programmable quantum computers available today~\cite{GOOGLE,IBM,Rigetti,Honeywell,IonQ,InnsbruckQC,UMDQC,MainzQC,YaleQC,BerkeleyQC}, computational instructions for quantum applications are typically
compiled into single- and two-qubit quantum gates~\cite{Qiskit,Rigetti_gateset,Maslov}.
At the physical, hardware-execution level, 
and in all current quantum-computer 
architectures, two-qubit gates
are about one to two orders of magnitude
more costly to implement 
than single-qubit gates \cite{TIQCReview,AWS_Rigetti}.
Therefore, improving two-qubit gate performance 
is critical to the utility and scalability of quantum computers. 

 For trapped ions, the best two-qubit gates are mediated by the harmonic motion through spin-dependent forces \cite{leibfried_2003b,MS-1,MS-2,MS-3}. For laser-based gates in multi-ion chains,  
a range of pulse-shaping protocols has been devised 
to decouple the multiple motional modes from the 
qubit degree of freedom, such as amplitude modulation 
\cite{zhu2006c,AM,UMDQC,AM3}, 
phase modulation \cite{PM}, 
frequency modulation \cite{FM}, 
or combinations thereof \cite{AMFM,Shapira}. 
These pulses can also create additional resilience to 
mode drift \cite{AMFM,Shapira} 
and gate-timing errors \cite{Shapira}, enable 
fast gate action \cite{Schaefer},  
or allow for simultaneous \cite{EASE, Figgatt2019} or multi-qubit \cite{Lu2019} gates. 

In \cite{AMFM} we reported a constructive method to calculate power-optimal pulse shapes based on the Mølmer-Sørensen protocol \cite{MS-1,MS-2,MS-3}. In practice, however, we expect a 
quantum gate to be imperfect due to limitations 
independent of the pulse shape, 
such as intensity fluctuations due to beam-power 
or beam-steering noise, 
motional-mode heating, motional dephasing, 
laser dephasing, off-resonant photon scattering, 
qubit dephasing or depolarization, and others.
Therefore, we need not require mathematical exactness
in constructing the pulse shape, but rather ensure that the error incurred by 
the imperfect pulse shape is much smaller than the 
other error mechanisms limiting the fidelity.
Based on this strategy we present here 
a pulse-shaping 
technique  
that includes a systematic method of 
trading negligible amounts of fidelity for 
power 
savings of up to an order of magnitude 
under realistic operating conditions for a trapped-ion quantum computer. Alternatively, 
by trading power savings for gate speed,
we are able to speed up two-qubit gates
for a given power budget.  We confirm this trade-off experimentally.
This method also formulates gates which are naturally robust to mode-frequency drifts.

\noindent 
{\it Protocol}.---Implementation 
of the trade-off strategy is based on 
a provably power-optimal pulse-shaping method 
for laser-based radial-mode gates \cite{AMFM}. Using simultaneous amplitude- and frequency-modulation, it produces a gate with a theoretical fidelity of 1 for any given gate time $\tau$ and includes a systematic method for actively stabilizing the fidelity with respect to motional-mode frequency drift. 
 We refer to this method as the 
exact AMFM method.
The laser-control pulses $g(t)$ are represented as Fourier 
integrals, 
$g(t)=\int_{-\infty}^{\infty} 
A(\omega) \exp(i\omega t) dt$, which are subsequently
discretized into a Fourier-sine series 
    $g(t) = \sum_{n=1}^{N_A} A_n 
    \sin(2\pi n t/\tau)$, 
where $N_A\sim 300$ achieves convergence
for typical two-qubit gates with 
$\tau\sim 100\,\mu$s. 
Thus, 
in contrast to previous methods that employ 
either a single \cite{IonQ,UMDQC} 
or a few \cite{Shapira} 
laser-frequency tones, 
our method uses a quasi-continuum of frequencies resulting 
in a chirped pulse 
of the form 
$g(t)=\Omega(t)\sin[\psi(t)]$, where 
$\psi(t)=\int_0^t \mu(t') dt'$ and  $\mu(t)$ is the detuning function. The resulting signal can be implemented straightforwardly with an arbitrary waveform generator.
Phase-space closure requires 
$\alpha_p^i=-\eta_p^i\int_0^{\tau} g(t) \exp(i\omega_p t)=0$, 
 for $i,p=1,\ldots,N$, 
where $p$ is the mode index, $i$ the ion index, $\eta_p^i$ the Lamb-Dicke parameter, $\omega_p$ the mode frequency, and $N$ the ion number.
Therefore, including 
stabilization against mode-frequency drift 
to an arbitrary order $K$, we 
require 
$\partial^K\alpha_p^i/\partial \omega_p^K=0$. 
This represents $Q=N(K+1)$ homogeneous, 
linear equations  
that, in matrix notation, may be written as 
$M\vec A^{\gamma}=0$, where $\gamma=1,\ldots,N_A-Q$ and
the set of nontrivial amplitude 
solutions, $\vec A^{\gamma}$, spans 
the null-space of the constraint matrix $M$. In general, the dimension $N_A$ of the 
frequency space is much larger than $Q$, which leaves a large null-space 
to optimize the gate power. Introducing the 
average power, 
$\bar P=(1/\tau)\int_0^{\tau}g^2(t)dt = 
(1/2)\sum_{n=1}^{N_A} A_n^2$, 
the gate angle 
$\chi_{ij}=\vec A^T V^{ij}\vec A$ 
is achieved with the minimum $\bar P$, 
if $\vec A$ is chosen as the eigenvector 
of $V^{ij}$ associated with the 
eigenvalue of 
largest absolute value. $V^{ij}$ is the null-space projected 
kernel matrix 
${\cal K}_{nm}^{ij}$ $=$ 
$\sum_p \eta_p^i\eta_p^j$ 
$\int_0^{\tau}dt_2\int_0^{t_2}dt_1 
\sin(2\pi n t_2/\tau)\sin(2\pi m t_1/\tau)$ 
$\sin[\omega_p(t_2-t_1)]$, which can be evaluated analytically since it contains only elementary functions. 
Since the 
ion number $N$ does not occur 
explicitly other than in the vector space size, the 
method is naturally 
scalable to any number of ions. 
Since the method involves only linear 
algebra, it is computationally 
efficient and straightforward to 
implement. 
Additional 
linear conditions, for instance 
stabilization against gate-timing 
errors~\cite{Shapira}, 
may be added at will. 
Thus the method also scales in 
the number and types of stabilization conditions. 
However, if more conditions are added, 
the size of the null-space contracts, and 
with it 
the number of accessible degrees of freedom, which leads to an increase in the power required.   
Conversely, for a set of constraints and a given power
budget, $\bar{P}$, there is a minimum gate-duration, $\tau_{\rm min}$, which roughly follows $\tau_{\rm min}\sim 1/\bar P$ 
for gate times $\tau > 100\,\mu$s. 
Thus, there is a trade-off 
between power requirement and gate duration. 

\begin{figure*}
\includegraphics[width=0.45\textwidth]{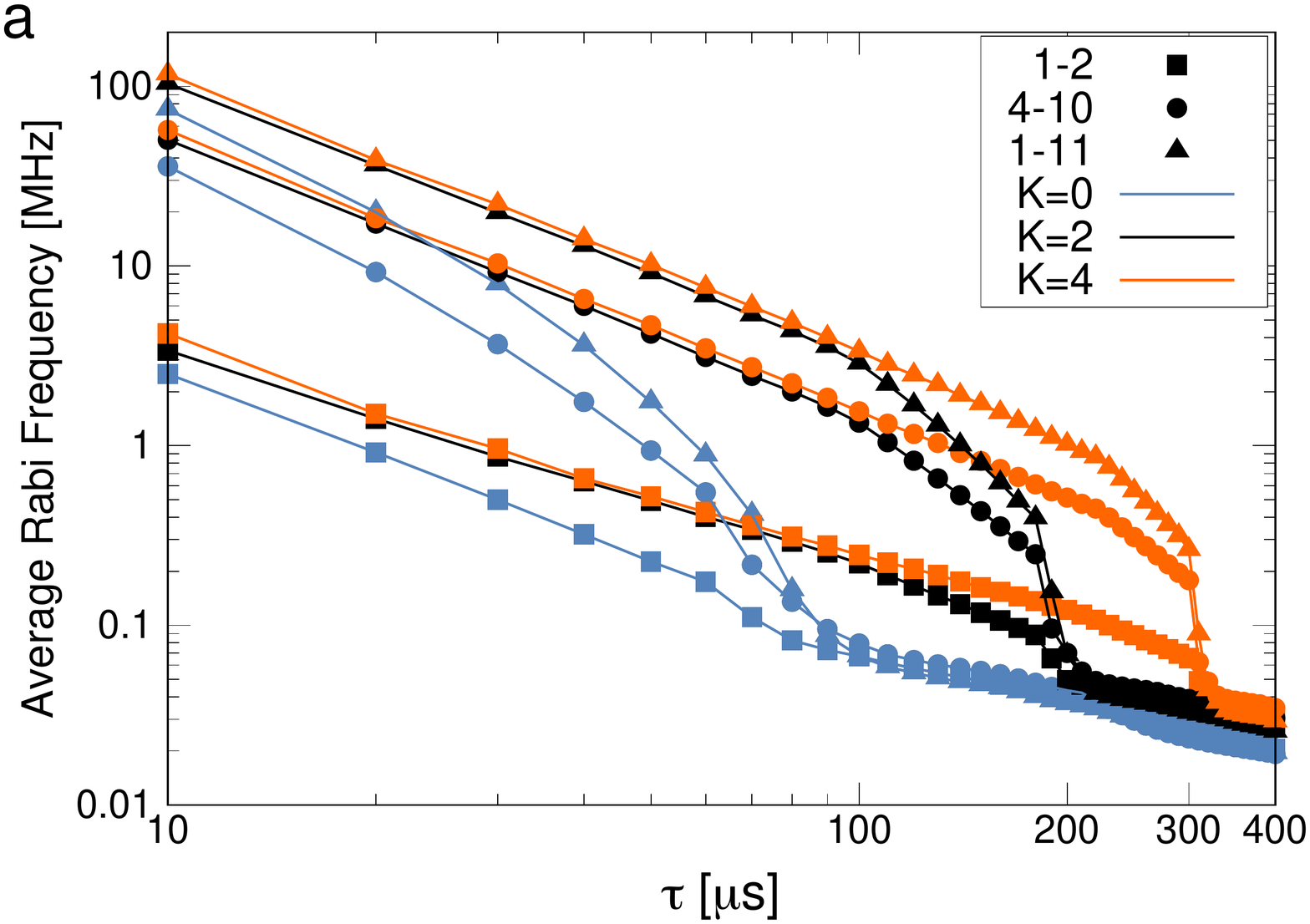}
\includegraphics[width=0.45\textwidth]{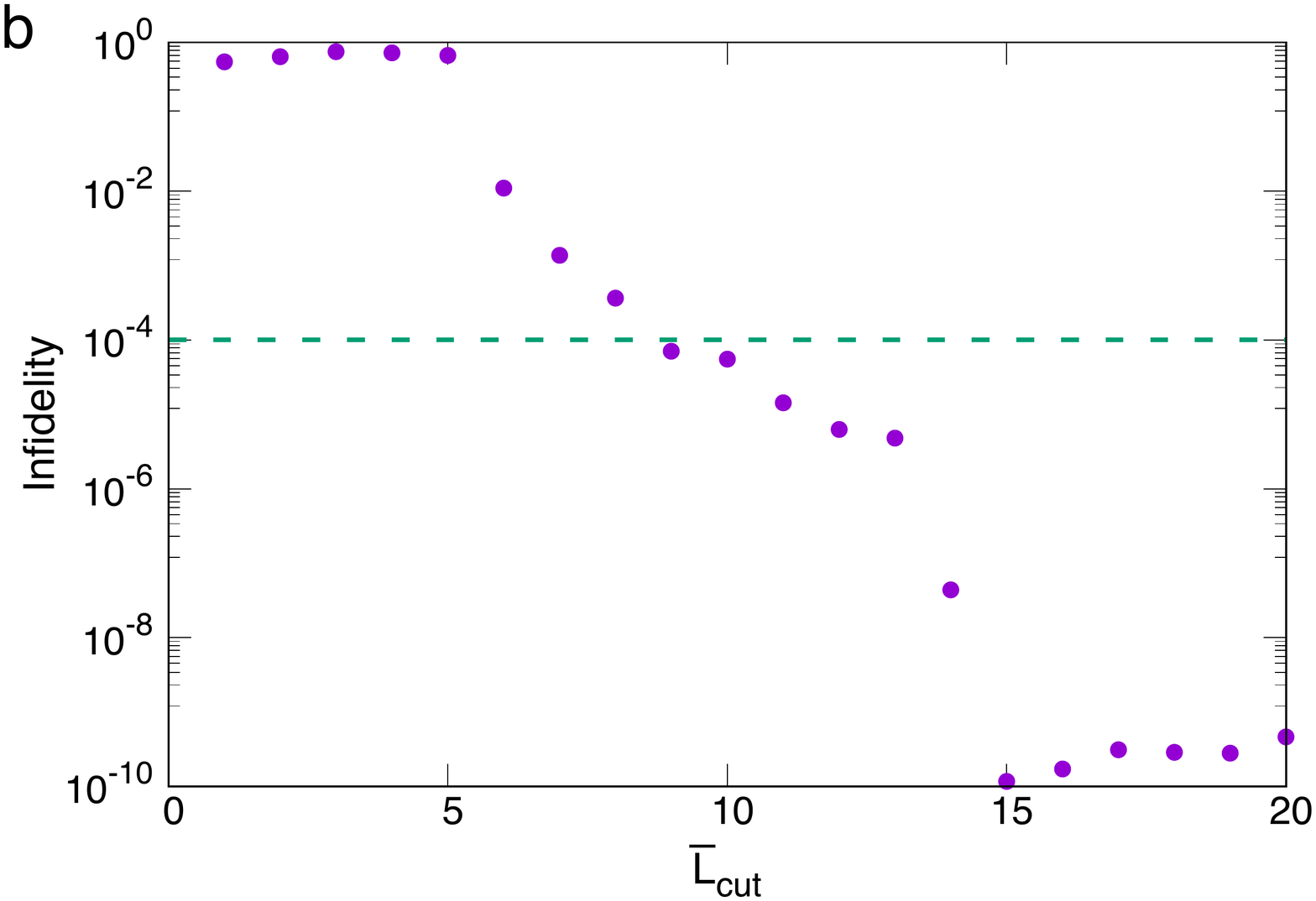}
\includegraphics[width=0.45\textwidth]{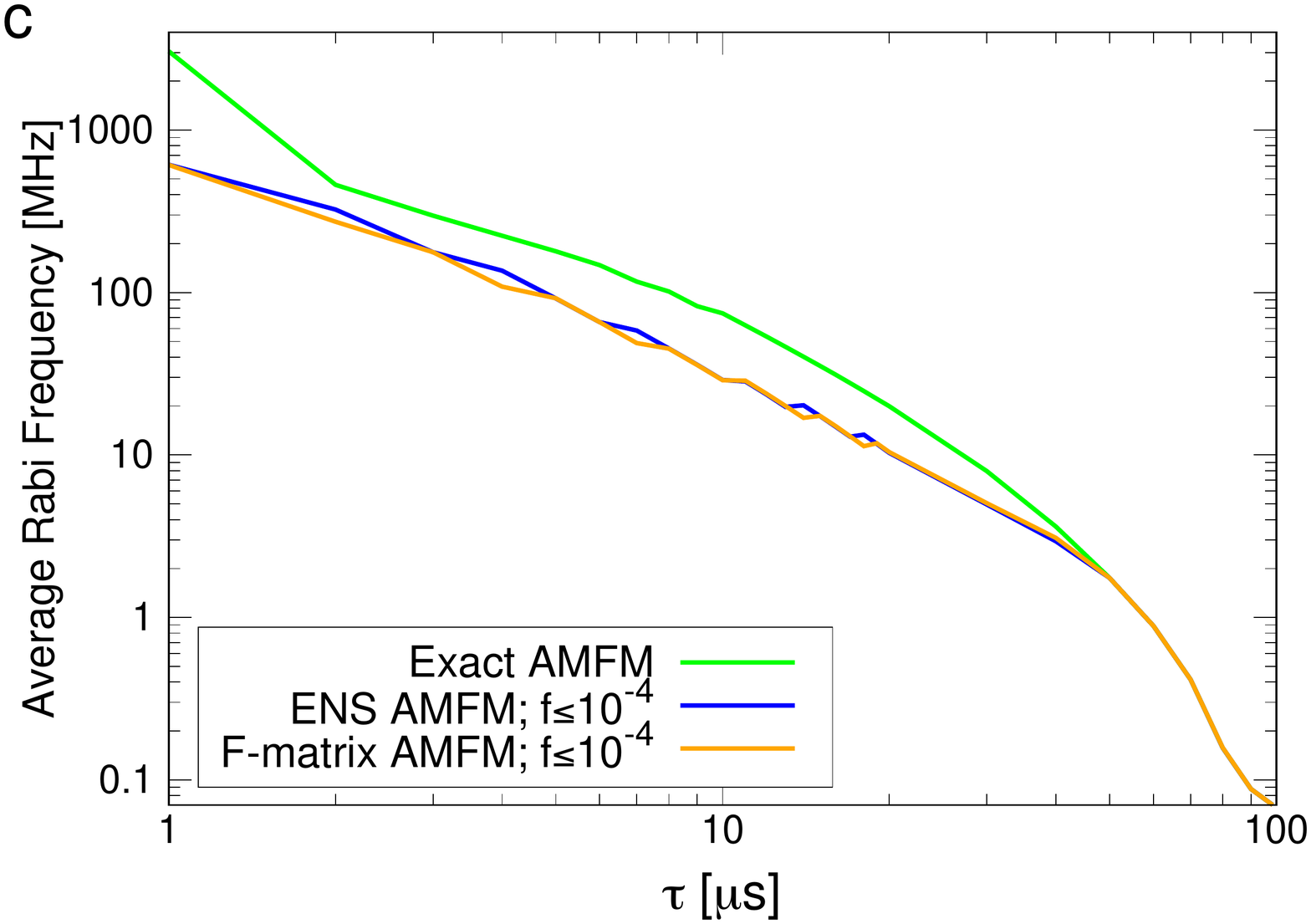}
\includegraphics[width=0.45\textwidth]{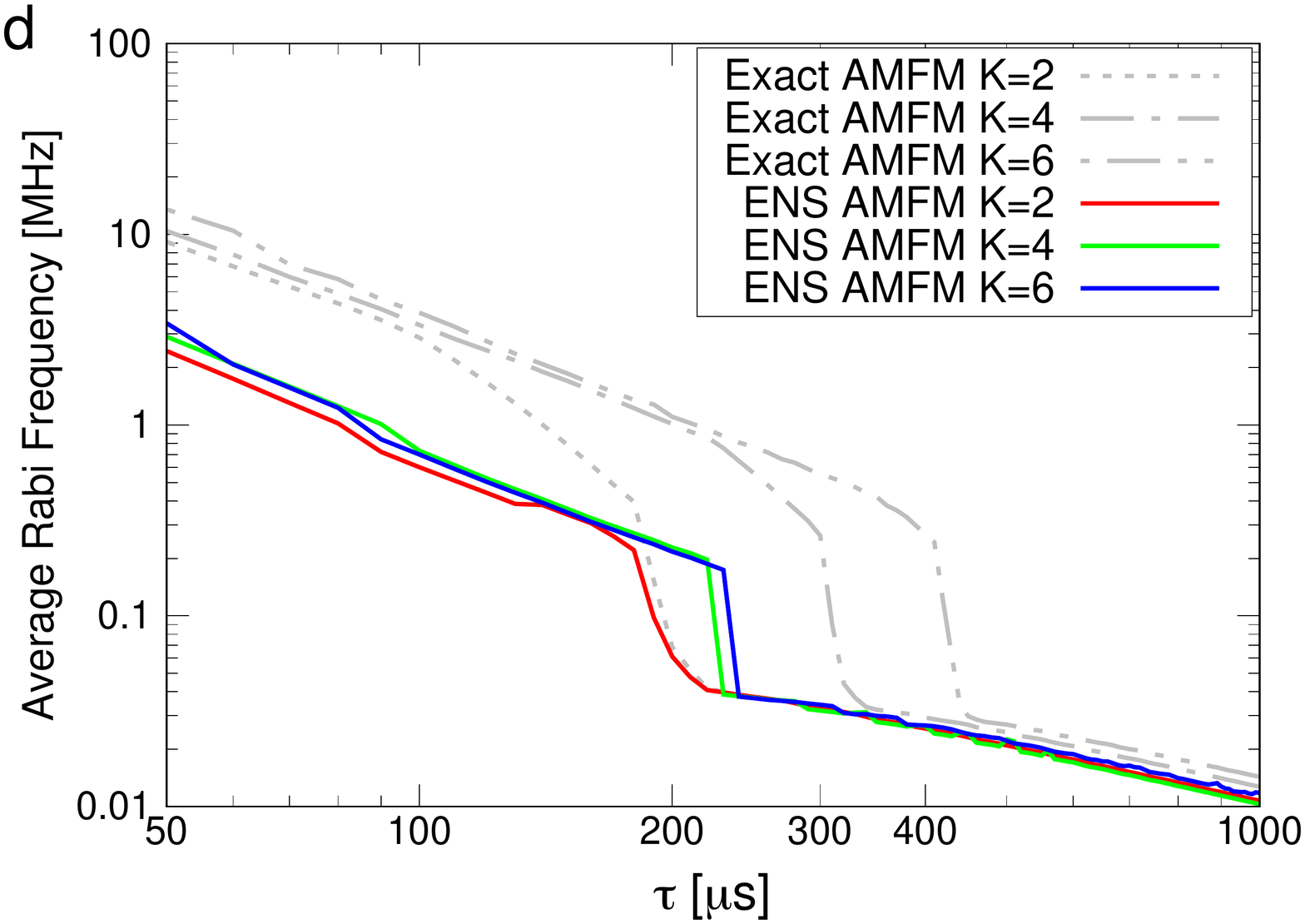}
\vspace{-2em}
\caption{\label{fig:protocol} Properties of gate pulses generated 
for a 15-ion, 11-qubit chain. 
(a) Power requirement of the exact AMFM gate 
as a function of gate time $\tau$  
for three different qubit pairs, $(1,2)$ (squares), 
$(4,10)$ (circles), and $(1,11)$ (triangles), and three 
different degrees of stability $K = 0$ (blue), $K=2$ (black), 
and $K=4$ (orange). 
(b) Infidelity $f$ as a function of $\bar L_{\rm cut}$
for $\tau=150\mu s$ gate pulses on qubits (1,11). 
A horizontal line at $f = 10^{-4}$, an acceptable amount 
of $f$ for contemporary experiments, was placed to guide the eye. 
(c) Three-way comparison of average power requirement between 
exact AMFM (green), 
$F$ matrix AMFM (orange), and ENS AMFM (blue), 
as a function of gate duration $\tau$ satisfying 
$f \leq 10^{-4}$. 
(d) Average power requirement as a function of $\tau$
for different degrees of stabilization $K$ for 
the ENS protocol (colored, solid lines) 
with $f \leq 10^{-4}$. 
Red, green, and blue lines are for 
$K=2$, $K=4$, and $K=6$, respectively. 
For comparison, the results for 
the exact AMFM for  
$K=2$ (dotted line) and $K=4$ (dot-dashed line) are 
copied from (a) without change, and $K=6$ (dot-dot-dashed line) is added to illustrate the trend. 
}
\end{figure*}

Figure~\ref{fig:protocol}(a) shows the power requirement
of the optimal, exact AMFM method   
for various qubit pairs ($i,j$) 
and degrees of stabilization $K$ as a function of
gate duration $\tau$ 
for a 15-ion chain, with an inter-ion spacing of $5\,\mu$m and the 11 central ions used as qubits.  
We see that the power requirement dramatically increases 
as we decrease $\tau$, 
exhibiting a step-like transition, 
whose location is nearly independent of 
the specific ion-pair. 
The reason is the following. To operate 
a two-qubit gate at low power, it is 
necessary that the basis frequencies,  
$2\pi n t/\tau$, have good overlap with 
the motional-mode frequencies, 
$\omega_p \lesssim 2\pi\times 3\,$MHz. 
As $\tau$ decreases, more 
basis frequencies are pushed out of 
the frequency range of 
the motional modes, effectively 
reducing the dimension of the null-space 
from which the power-optimal solutions 
are drawn. The step in power results when we 
run out of null-space vectors with 
good motional-mode frequency overlap. 
Since increasing $K$ reduces 
the null-space dimension even further, 
the power step happens at larger 
gate times $\tau$ for larger $K$. 
Similar to \cite{Schaefer}, our scheme cancels carrier excitations to first order due to the sinusoidal nature of our basis functions. While Fig. \ref{fig:protocol} shows gates $<100\mu$s, which remain approximately in the Lamb-Dicke regime, further investigation is needed to ensure the standard MS formalism with its perturbative expansions of the Hamiltonian is still valid in this short-pulse regime. 

The two-qubit gate infidelity $f$ at zero 
temperature is given by 
$ f = 0.8 \vec{A}\,^T F \vec{A}$,
where 
$ F_{nm} = \sum_{p=1}^N [(\eta_p^i)^2+(\eta_p^j)^2] 
\sigma_{np}\sigma_{mp}$
is the infidelity matrix  
and
$ \sigma_{np} = \int_0^\tau \sin(2\pi n t/\tau) \exp(i\omega_p t) dt$ \cite{GREEN}.
The infidelity $f$ may be systematically 
controlled by expanding  
the gate-pulse amplitudes 
$\vec{A}$ in the set of 
eigenvectors $\vec W_l$ of $F$ that correspond to its smallest absolute eigenvalues up to
$l= L_{\rm cut}$, i.e.  
$\vec A=\sum_{l=1}^{L_{\rm cut}}B_l\vec W_l$. 
The gate angle is then given by 
$\chi_{ij}=\vec A^T {\cal K}^{ij}\vec A=
\vec B^T S^{ij}\vec B$, where 
$S_{ll'}^{ij}=\vec W_l^T {\cal K}^{ij} \vec W_{l'}$. 
The coefficients $\hat{B}_l$ that minimize the 
power requirement are then the components of the 
eigenvector of $S^{ij}$ with the largest-modulus eigenvalue.
We call this protocol $F$-matrix AMFM (see Supplemental Material Sec.~\ref{FinexactAMFM} for expanded mathematical details). 

In Fig.~\ref{fig:protocol}(b) we show the infidelity computed 
according to 
$F$-matrix AMFM 
as a function of 
$\bar L_{\rm cut}=N_A - L_{\rm cut}$, 
the complement of $L_{\rm cut}$
with respect to $N_A$. 
We observe that the infidelity rapidly decreases with increasing $\bar L_{\rm cut}$, 
reaching $10^{-4}$ at $\bar L_{\rm cut} \geq 9$. This infidelity is deemed acceptable in contemporary experiments,
given that it roughly corresponds to the 
spontaneous scattering limit \cite{spontaneous, GREEN}. 
This shows that a large number of eigenvectors of $F$ 
may indeed be used as the variational 
space for minimizing the power 
with only a negligible fidelity cost.

In Fig.~\ref{fig:protocol}(c) we compare the power requirement of $F$-matrix AMFM 
with that of the exact AMFM as a function of gate duration $\tau$. 
We see that $F$-matrix AMFM provides a power advantage for $\tau \lesssim 50\,\mu$s for qubit pair $(1,11)$ in a 15-ion, 11-qubit chain.

An important feature of the exact AMFM  
is its ability to 
actively introduce robustness against experimental 
parameter drift and fluctuations. 
While, in principle, the $F$-matrix approach 
can be stabilized as well, 
the following approach is more 
straightforward to 
implement.

\begin{figure*}[t!]
\includegraphics[width=0.45\textwidth]{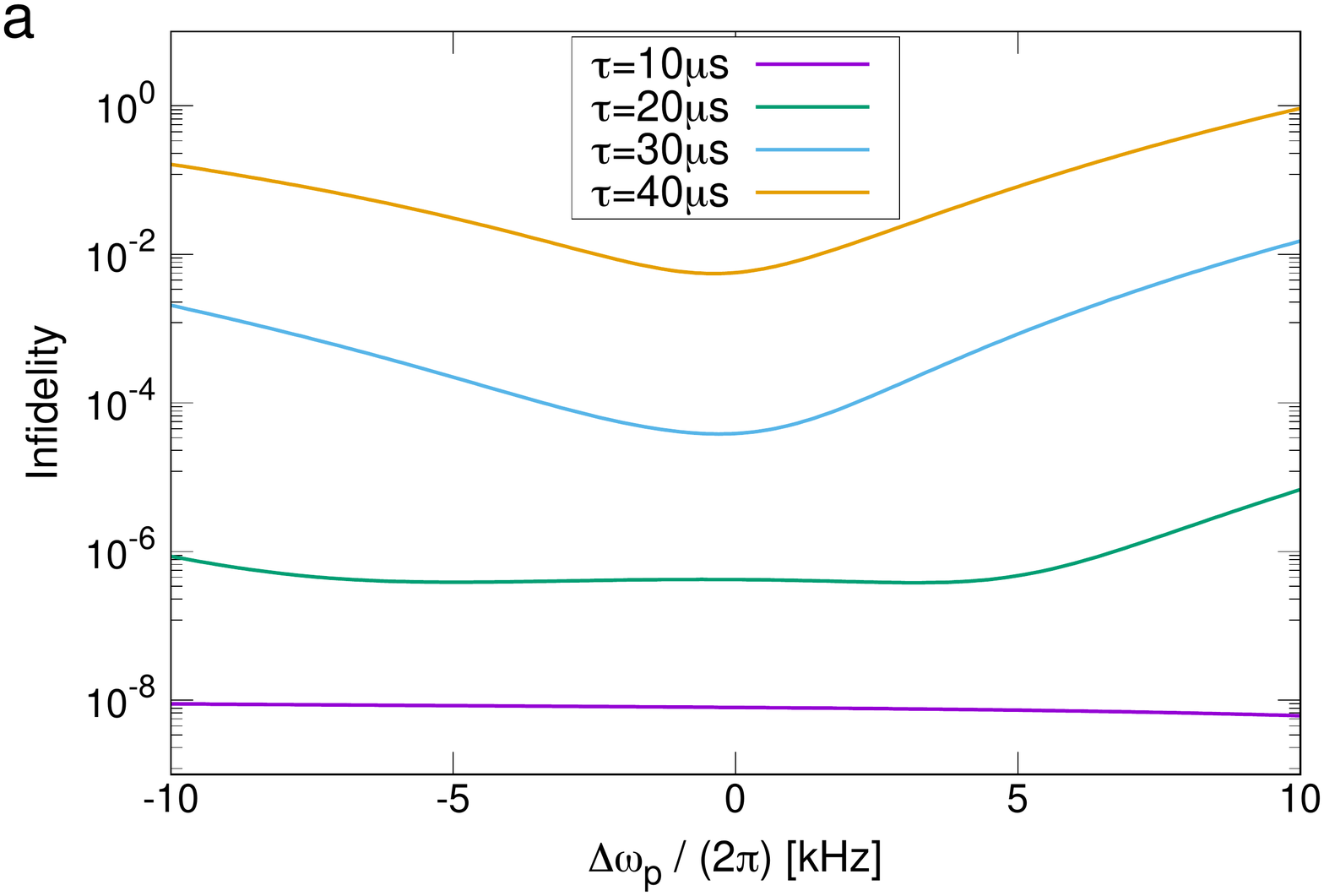}
\includegraphics[width=0.45\textwidth]{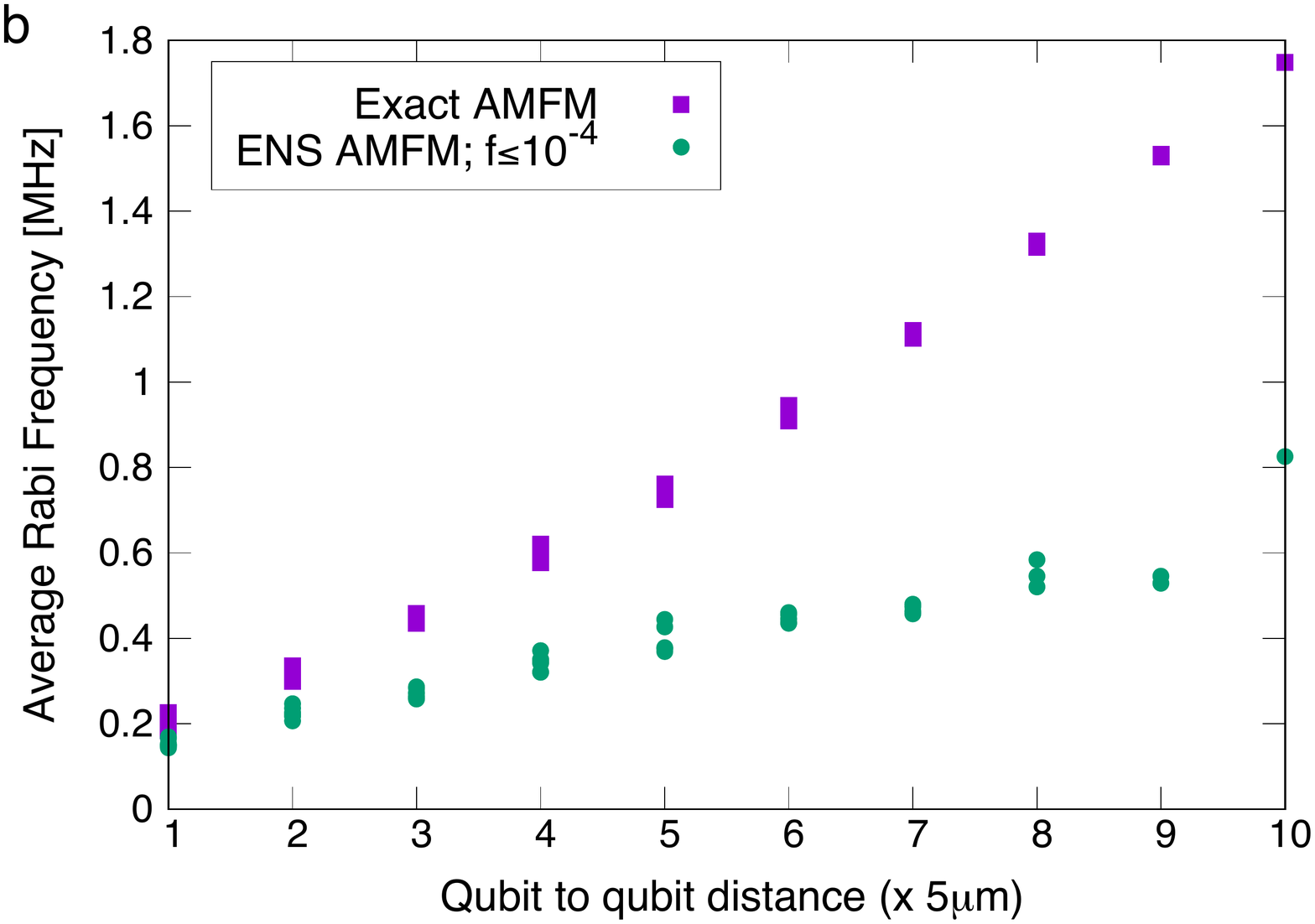}
\vspace{-2em}
\caption{\label{fig:disc} Trade space 
for a 15-ion, 11-qubit chain. 
(a) Infidelity $f$ as a function of uniform mode-frequency 
drift $\omega_p\rightarrow \omega_p + \Delta \omega_p$ 
on qubits $(1,11)$ 
($\bar L_{\rm cut}=12$) 
for four different gate times $\tau$.
(b) Average power requirement as a function of distance between the 
qubits for a $\tau$=50\,$\mu$s gate.
Purple squares: exact AMFM; green circles: 
$F$-matrix protocol. 
}
\vspace{-1em}
\end{figure*}

We can
relax the stringent requirement of 
perfect decoupling between the qubit and motional states as an alternative way to 
introduce a negligible amount of infidelity. 
According to the exact AMFM protocol this may 
be accomplished by 
constructing an approximate null space of $M$ that 
now also includes eigenvectors with non-zero 
eigenvalues, as long as their moduli 
are smaller than a 
pre-determined, non-zero value. The Supplemental Material Sec.~\ref{sec:Sliding} contains expanded mathematical detail.
Figure~\ref{fig:protocol}(c) shows that 
the power requirement of this extended null-space (ENS) approach 
is nearly identical to the 
power requirement of the 
$F$-matrix approach. 
In Fig.~\ref{fig:protocol}(d) we compare the pulse-power 
requirements of stabilized pulses 
produced according to the exact AMFM method 
and the ENS method, for which implementation of the stability conditions is straightforward. 
Over a large span of gate times, 
the ENS method offers significant power savings
for stabilized pulses. In particular, 
we find that for stability degree $K = 6$ 
and gate duration $\tau = 250\,\mu$s, 
the power saving can be as large as a factor of 15. 

\begin{figure}
\includegraphics[width=0.48\textwidth]{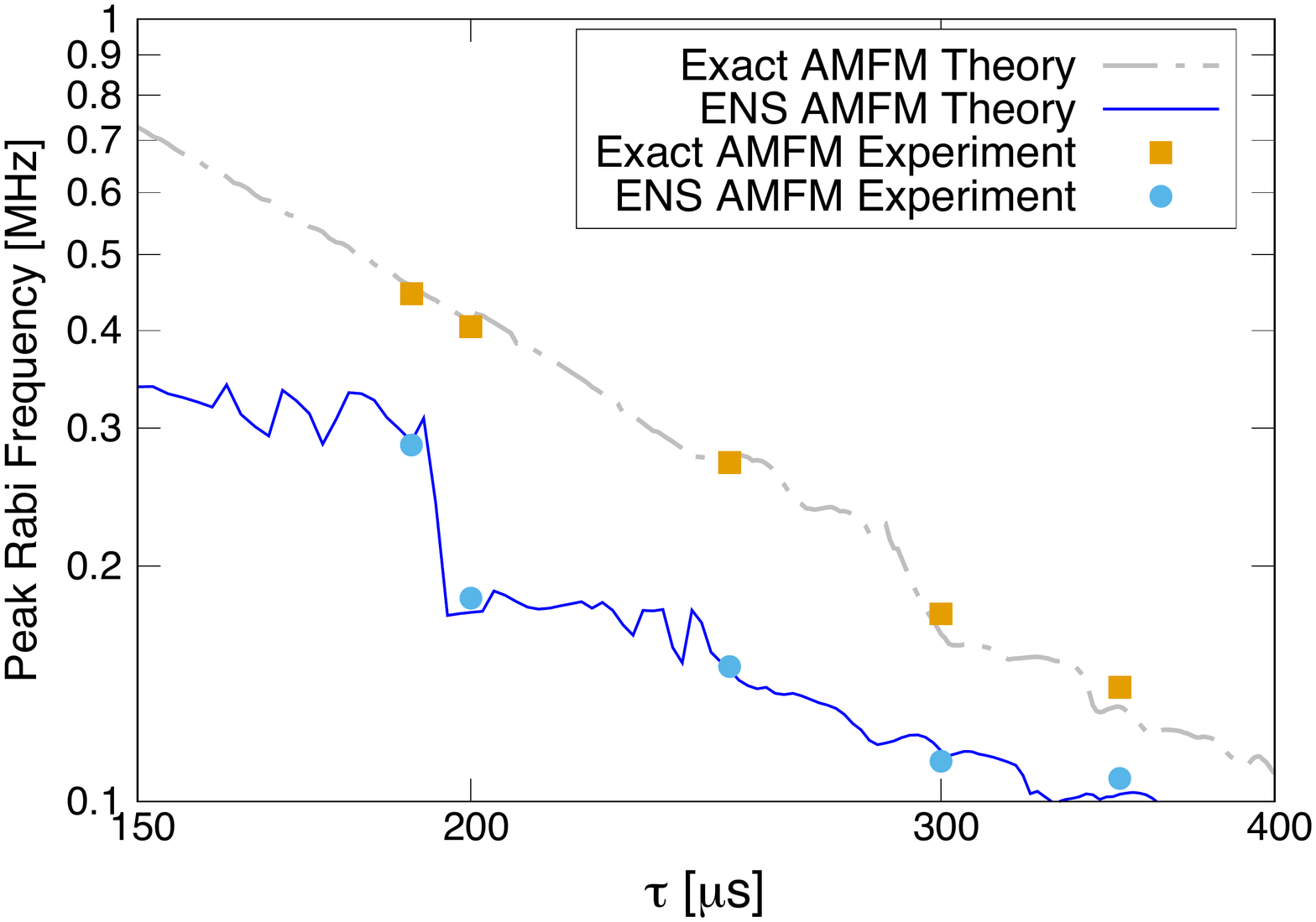}
\vspace{-3em}
\caption{\label{fig:exp} 
Comparison of 
peak power 
requirements of ENS gates (blue solid line)
with exact AMFM gates (grey dashed line) 
on qubits $(4,5)$ as a function of 
gate time 
$\tau$ ($K=4$)  
for 
$f\leq10^{-4}$ on a seven-ion, 
five-qubit chain. 
Experimental results for different gate times,
implementing 
pulses constructed according 
to the exact AMFM and ENS protocols,  
are shown as orange squares and 
blue circles, respectively. 
The experimental error bars are smaller than the plot symbols.}
\vspace{-1em}
\end{figure}

We note that faster gates 
come with exponentially decreasing infidelity and increased natural stability against mode-frequency drift, even in the absence 
of active stabilization. We illustrate this 
in Fig.~\ref{fig:disc}(a), which shows 
that increasing the gate speed from 
$40\,\mu$s to $10\,\mu$s reduces the 
infidelity by about 6 orders of magnitude, 
reaching below $f=10^{-8}$ over a drift-frequency 
range larger than $\pm 10\,$kHz at 
$\tau=10\,\mu$s. While pulses this short may not be practical, the stability they provide can be propagated to longer gates at the cost of power optimality by reducing the power and repeating the pulse sequence multiple times.

According to 
Fig.~\ref{fig:protocol}(a)
qubit pairs 
that are farther apart from each other require more power. 
Therefore, given 
a fixed power budget, 
instead of trading the power savings afforded by our pulse 
design for gate duration, we can alternatively 
trade the savings for better qubit connectivity. This power-connectivity trade-off can play a critical role in harnessing the power of quantum computation since matching hardware and application connectivity is crucial for performance in a future quantum operating system \cite{ar:LinkePNAS}.
In Fig.~\ref{fig:disc}(b) we show the power requirement for $\bar L_{\rm cut} = 12$ and $\tau = 50\,\mu$s as a function of qubit 
distance for the exact AMFM and the $F$-matrix AMFM. 
We see that the $F$-matrix AMFM requires 
factors of about 2 to 4 smaller power when compared to the exact AMFM counterparts.

\noindent 
{\it Experiment}.---
We implement the exact AMFM and ENS pulses on a 
programmable, fully-connected trapped-ion quantum computer at the University of Maryland \cite{Landsman2019}. 
We trap seven ${}^{171}{\rm Yb}^+$ ions in 
a 1D chain and use the middle five as qubits. 
The ions are laser-cooled close to the motional ground state, and 
then optically pumped to $\ket{0}$. Coherent operations are driven by a pair of counter-propagating Raman beams, 
one of which is split into individual addressing beams, each focused on one ion. These beams are controlled independently 
by RF pulses generated by arbitrary waveform generators (AWGs), which enable the implementation of pulse shapes from a broad range of frequencies, amplitudes, and phases. 

Figure~\ref{fig:exp} shows the theoretically predicted and experimentally measured power requirements as a function of gate duration for 
exact AMFM and ENS-based pulses. 
We chose infidelity $f = 10^{-4}$ to determine the 
ENS since the gates are limited to $10^{-2}$ 
by other imperfections. We chose to stabilize these pulses to degree $K = 4$.

Due to the limited amplitude resolution of the AWG (14-bit DAC), the relative amplitudes of basis frequencies smaller than $10^{-4}$ are neglected in the experimental implementation. We confirm numerically that this does not significantly impact the theoretical fidelity and stability of the resulting gates. 
We verify a successful implementation of a pulse by observing a continuous coherent transfer of population between the $\ket{00}$ and $\ket{11}$ states when the applied laser power is varied. We then calculate the minimal Rabi frequency $\Omega_0 =s\textbf{}\Omega_{\max}$ needed to perform a maximally entangling gate, where $s \leq 1$ is a scale factor and $\Omega_{\max}$ is the maximum Rabi rate available. 
We further verify the creation of the maximally entangled state by measuring the parity contrast for some of the pulses \cite{UMDQC}. 

All the experimentally determined Rabi frequencies for the
two-qubit gates using either the ENS-
or the exact AMFM protocols fall within $\pm 10\%$ of  
the respective theoretically predicted values.
The small discrepancies between experimental values and
theory predictions are due to uncertainties
in the Lamb-Dicke parameters and mode frequencies (see Supplemental Materials \ref{sec:ImpDet}). 

Finally, we verify the robustness of the gate solutions against drifts in  the motional modes. We compare the phase space loop closure of an ENS gate stabilized to two different degrees, $K=1$ and $K=5$. The results, described in the Supplemental Material \ref{sec:StabPulse}, are consistent with predictions.
   
\smallskip\noindent 
{\it Outlook}.---
The impossible trinity formed by power, gate duration, and fidelity discussed and illustrated in this paper has 
an analog in the 
chip design community, where 
there are 
well-known 
fundamental trade-offs between 
power, performance, and area, 
commonly referred to as PPA 
(see, e.g., \cite{PPA}). 
Given that the physical-level infidelity affects, for instance, the cost of implementing 
quantum error correction to achieve a target logical-level infidelity, our investigations could be considered a quantum version of PPA trade-offs. Given the enormous utility and impact of a careful PPA study for system-on-a-chip design, our results may contribute to future quantum processor optimization. We are convinced that this holistic optimization of all facets of the design based on the PPA trade-off is another stepping stone 
toward successful practical quantum computing. 

{\it Acknowledgements}.---
We thank Cinthia Huerta Alderete and Yingyue Zhu for help with the experiment. N.M.L. acknowledges financial support from the NSF grant no. PHY-1430094 to the PFC@JQI, and the Maryland-ARL Quantum Partnership, grant no. W911NF1920181. A.M.G. is supported by a JQI Postdoctoral Fellowship.

\bibliographystyle{unsrt}
\bibliography{fast-ms-gate.bib}

\newpage

\section*{Supplemental Material}
\renewcommand\theequation{S\arabic{equation}}    
\renewcommand\thetable{S\arabic{table}}    
\renewcommand\thefigure{S\arabic{figure}}   
\renewcommand\thesubsection{S\arabic{subsection}}
\renewcommand\thesection{S\arabic{section}}
\setcounter{section}{0}    
\setcounter{subsection}{0}    
\setcounter{equation}{0}    
\setcounter{table}{0}    
\setcounter{figure}{0}    
\comment{
\section{Exact AMFM}
\label{sec:Exact}

We summarize in this section
the exact AMFM method described in \cite{AMFM}.
To start, we expand the laser pulse shape $g(t)$
as a Fourier-sine series
\begin{align}
    \label{prior1}
    g(t) = \sum_{n=1}^{N_A} A_n \text{sin}\Big(\frac{2\pi n}{\tau}t\Big), 
\end{align}
where $A_n$, $n=1,\ldots,N_A$, are real expansion amplitudes,
$N_A$ is chosen large enough to achieve convergence, 
and $\tau$ is the gate time.
We require the pulse shape to satisfy 
the stabilized phase-space closure conditions
$\alpha_p^i(\tau)$, i.e., 
\begin{equation}
\bigg(\frac{\partial^k}{\partial \omega_p^k}\bigg) \alpha_p^i(\tau) 
= - \bigg(\frac{\partial^k}{\partial \omega_p^k}\bigg) 
\eta_p^i \int_0^{\tau} g(t) e^{i\omega_p t} dt = 0
\label{prior2}
\end{equation}
for $k=0,1,\ldots,K$, where $K$ is the 
desired degree of stabilization with respect 
to the motional-mode 
frequencies $\omega_p$, $p=1,\ldots,N$, 
$i=1,\ldots,N$ is the ion-number index,  
$N$ is the number of ions, 
$g(t)$ 
is the pulse function defined in (\ref{prior1}), and 
$\eta_p^i$ is the Lamb-Dicke parameter
\cite{LD}, which describes the coupling strength of ion 
number $i$ to motional-mode number $p$. 
Using the form of $g(t)$ from (\ref{prior1}) in (\ref{prior2}), we obtain
\begin{align}
\label{prior3}
& - \bigg(\frac{\partial^k}{\partial \omega_p^k}\bigg) \eta_p^i 
\sum_{n=1}^{N_A} A_n \int_0^{\tau} \text{sin}\Big(\frac{2\pi n}{\tau}t\Big) e^{i\omega_p t} dt & \nonumber \\
&= \sum_{n=1}^{N_A} M_{mn}A_n =  M\vec{A} = 0,
\end{align}
where  
$m=p+Nk$ is the linearized index of  
the $\alpha_p^i(\tau)$ equations 
for different $p$ and $k$ [$m = 1, 2, \ldots, N(K+1)$], and 
\begin{align}
    \label{prior4}
    M_{mn} = - \bigg(\frac{\partial^k}{\partial \omega_p^k}\bigg) 
    \eta_p^i \int_0^{\tau} \text{sin}\Big(\frac{2\pi n}{\tau}t\Big) e^{i\omega_p t} dt.
\end{align}
The aim is to find the null space of the matrix $M$, 
i.e., the set of $N_0=N_A-Q$ linearly independent, nontrivial  
vectors $\vec{A}^{(\alpha)}$, $\alpha=1,\ldots,N_0$, that 
satisfy $M\vec{A}^{(\alpha)}=0$. 
These vectors form a solution subspace from 
which the optimal solution vector
\begin{align}
    \hat{\vec{A}}=\sum_{\alpha=1}^{N_0} \Lambda_{\alpha} \vec{A}^{(\alpha)}
\end{align} 
is chosen, following the 
power-optimization procedure described below. 
In order for (\ref{prior3}) to have non-trivial solutions, 
we require that $N_A>N(K+1)$. 
Thus, in general, the rank-deficient, 
rectangular matrix $M$ has $N(K+1)$ 
non-zero eigenvalues and $N_0 = N_A-N(K+1)$ 
zero eigenvalues.  This 
suggests to 
multiply (\ref{prior3}) from the left with the transpose, 
$M^T$, of $M$, 
which turns (\ref{prior3}) into 
\begin{equation}
\Gamma \vec A = 0, 
\label{prior5} 
\end{equation}
where 
\begin{equation} 
\Gamma=M^T M
\label{Gamma}
\end{equation} 
is a symmetric matrix. 
Section~S20 of Ref.~\cite{AMFM} shows that the operation of 
multiplication with $M^T$ does not change the structure of 
the solution space. 
Thus, (\ref{prior3}) is now 
turned into an eigenvalue problem with the symmetric 
matrix $\Gamma$, and we are looking for 
the $N_0$ eigenvectors $\vec A^{(\alpha)}$ 
of $\Gamma$ with eigenvalues 0. The $N_0$ nontrivial vectors 
$\vec A^{(\alpha)}$ with eigenvalues 0 span the kernel of 
the matrix $\Gamma$, also known as the 
{\it null space} of $\Gamma$.
Since all null-space vectors $\vec A^{(\alpha)}$ have the common 
eigenvalue 0, the null space is (highly) degenerate. Thus, 
any linear combination of the $N_0$ null-space vectors 
$\vec A^{(\alpha)}$ are also null-space vectors, and we may 
assume that the $\vec A^{(\alpha)}$ form an orthonormal basis 
of the null space according to 
\begin{equation}
\vec A^{(\alpha)}\,^T \vec A^{(\beta)} 
= \delta_{\alpha\beta}, 
\label{prior6}
\end{equation}
where $\delta_{\alpha\beta}$ is the Kronecker symbol. 
If the pulse acts 
simultaneously on ions $i$ and $j$, 
the gate angle $\varphi_{ij}$ of a 
Mølmer-Sørensen (MS) 
gate is given by \cite{AMFM}
\begin{equation}
\varphi_{ij} = 2(\chi_{ij} + \chi_{ji}), 
\label{prior7}
\end{equation}
where 
\begin{equation}
\label{prior8}
\chi_{ij} = \sum_{p=1}^N\, \eta_p^i \eta_p^j 
\int_0^{\tau} dt_2 \int_0^{t_2} dt_1
g(t_2) g(t_1)\sin[\omega_p(t_2-t_1)] . 
\end{equation}
A maximally entangling gate is 
achieved for $\varphi_{ij}=\pm\pi/2$. According to 
(\ref{prior8}), $\chi_{ij}=\chi_{ji}$, i.e., 
$\varphi_{ij}=4\chi_{ij}$, so that 
a maximally entangling gate requires 
\begin{equation}
|\chi_{ij}| = \frac{\pi}{8}. 
\label{prior8a}
\end{equation}
Our goal now is to 
linearly combine the orthonormal null-space vectors 
$\vec A^{(\alpha)}$ with real expansion amplitudes 
$\Lambda_{\alpha}$ 
to find 
the optimal null-space vector 
\begin{equation}
\hat{\vec A} = \sum_{\alpha=1}^{N_0} \Lambda_{\alpha} 
\vec A^{(\alpha)} 
\label{prior9}
\end{equation}
such that
\begin{equation}
\hat g(t) 
= \sum_{n=1}^{N_A} \hat A_n 
\sin \left(2\pi n \frac{t}{\tau} \right)
\label{prior10}
\end{equation}
is optimal in the sense that it produces 
$|\chi_{ij}|=\pi/8$, according to (\ref{prior8a}), and 
has the smallest possible norm 
\begin{equation}
\gamma^2 = 
|| 
\hat g(t) 
||^2 = 
\frac{2}{\tau} 
\int_0^{\tau} 
\left[ 
\hat g(t) 
\right]^2 \, dt 
= \min_{\Lambda_{\alpha}} \sum_{n=1}^{N_A} 
\hat A_n^2, 
\label{prior11}
\end{equation}
which entails the smallest possible average 
power needed to execute a maximally entangling MS gate. 
The average pulse power $\bar{P}$ is defined as 
\begin{align}
\label{prior11a}    
    \bar{P} = \Bigg[\frac{1}{\tau}\int_0^{\tau} g^2(t) dt \Bigg]^{1/2}.
\end{align}
Using (\ref{prior10}) and (\ref{prior8a}) in 
(\ref{prior8}), we obtain 
\begin{align} 
\frac{\pi}{8} &= \Bigg|
\sum_{p=1}^N \eta_p^i \eta_p^j 
\int_0^{\tau}\, dt_2\, \int_0^{t_2}\, dt_1 \, 
\nonumber\\
& \qquad
\hat g(t_2)\, \hat g(t_1)\, 
\sin\left[\omega_p(t_2-t_1)\right] \Bigg|
\nonumber\\
&= \left| \hat{\vec A\,}^T D \hat{\vec A} \right| , 
\label{prior12}
\end{align}
where $D$ is a real $N_A\times N_A$ matrix with matrix elements 
\begin{align}
D_{nm} &= \sum_{p=1}^N \eta_p^i \eta_p^j 
\int_0^{\tau}\, dt_2\, \int_0^{t_2}\, dt_1 \, 
\nonumber\\
&\sin \left(2\pi n \frac{t_2}{\tau} \right)
\sin\left[\omega_p(t_2-t_1)\right] 
\sin \left(2\pi m \frac{t_1}{\tau} \right) . 
\label{prior13}
\end{align} 
Since $\hat{\vec A}^{\;T} D \hat{\vec A}$ is a scalar, 
we can also write 
\begin{equation}
\hat{\vec A}^{\;T} D \hat{\vec A} = \frac{1}{2} \left[
\hat{\vec A}^{\;T} D \hat{\vec A} + 
\left(\hat{\vec A}^{\;T} D \hat{\vec A}\right)^T \right] 
= \hat{\vec A}^{\;T} S \hat{\vec A} , 
\label{prior14}
\end{equation} 
where 
\begin{equation}
S = \frac{1}{2} \left[ D + D^T \right] 
\label{prior15}
\end{equation} 
is a symmetric matrix. 
Using (\ref{prior15}) and (\ref{prior14}) in 
(\ref{prior12}) we obtain 
\begin{equation}
\frac{\pi}{8} = \left|
{\vec \Lambda}^T R \vec \Lambda 
\right| , 
\label{prior15a}
\end{equation} 
where $\vec\Lambda$ is the vector of expansion 
amplitudes $\Lambda_{\alpha}$, 
$\alpha=1,\ldots,N_0$, and 
$R$ is the symmetric, reduced $N_0\times N_0$ matrix 
with matrix elements 
\begin{equation}
R_{\alpha\beta} = {\vec A}^{(\alpha)}\,^T S 
{\vec A}^{(\beta)},\ \ \ 
\alpha, \beta=1,\ldots,N_0. 
\label{prior16}
\end{equation} 
Since $R$ is symmetric, it can be diagonalized, 
\begin{equation}
R\, \vec V^{(k)} = \lambda_k\, \vec V^{(k)} , \ \ \ 
k=1,\ldots,N_0 , 
\label{prior16a}
\end{equation} 
where, since $R$ is real and symmetric, the 
eigenvectors $\vec V^{(k)}$ can be assumed orthonormal. 
We now linearly combine the vector of 
expansion amplitudes $\vec \Lambda$ from the 
set of vectors $\vec V^{(k)}$ according to 
\begin{equation}
\vec\Lambda = \sum_{k=1}^{N_0} v_k \vec V^{(k)}. 
\label{prior17}
\end{equation} 
According to (\ref{prior11}), 
we now have to determine the expansion 
amplitudes $v_k$ such that 
\begin{equation}
\gamma^2 = \min_{v_k} \hat{\vec A}^{\;T} \hat{\vec A} = 
\min_{v_k} {\vec \Lambda}^T {\vec \Lambda} = 
\min_{v_k} \sum_{k=1}^{N_0} v_k^2 
\label{prior18}
\end{equation} 
under the condition 
\begin{equation}
\frac{\pi}{8} = | {\vec \Lambda}^T R {\vec \Lambda} | = 
|\sum_{k=1}^{N_0} v_k^2 \lambda_k|. 
\label{prior19}
\end{equation} 
Thus, our optimization problem 
is solved: The optimal pulse (\ref{prior10}) is constructed 
with the help of the amplitudes 
\begin{equation}
\hat {\vec A} = \sum_{\alpha=1}^{N_0} \Lambda_{\alpha}^{(k_{\rm max})}  
\vec A^{(\alpha)}, 
\label{prior20}
\end{equation} 
where $k_{\rm max}$ is the index of the 
eigenvalue $\lambda_k$ of (\ref{prior16a}) with the largest modulus 
$|\lambda_k|$, and 
\begin{equation}
\vec \Lambda^{(k_{\rm max})} = v_{k_{\rm max}} \vec V^{(k_{\rm max})} , 
\label{prior21}
\end{equation} 
where 
\begin{equation}
v_{k_{\rm max}} = 
\left(\frac{\pi}{8 |\lambda_{k_{\rm max}}|}\right)^{1/2} . 
\label{prior22}
\end{equation} 
 }

\section{F-matrix protocol}
\label{FinexactAMFM}
To start, we expand the laser pulse shape $g(t)$
as a Fourier-sine series
\begin{align}
    \label{prior1}
    g(t) = \sum_{n=1}^{N_A} A_n \text{sin}\Big(\frac{2\pi n}{\tau}t\Big), 
\end{align}
where $A_n$, $n=1,\ldots,N_A$, are real expansion amplitudes,
$N_A$ is chosen large enough to achieve convergence, 
and $\tau$ is the gate time.
We require the pulse shape to satisfy 
the stabilized phase-space closure conditions
$\alpha_p^i(\tau)$, i.e., 
\begin{equation}
\bigg(\frac{\partial^k}{\partial \omega_p^k}\bigg) \alpha_p^i(\tau) 
= - \bigg(\frac{\partial^k}{\partial \omega_p^k}\bigg) 
\eta_p^i \int_0^{\tau} g(t) e^{i\omega_p t} dt = 0
\label{prior2}
\end{equation}
for $k=0,1,\ldots,K$, where $K$ is the 
desired degree of stabilization with respect 
to the motional-mode 
frequencies $\omega_p$, $p=1,\ldots,N$, 
$i=1,\ldots,N$ is the ion-number index,  
$N$ is the number of ions, 
$g(t)$ 
is the pulse function defined in (\ref{prior1}), and 
$\eta_p^i$ is the Lamb-Dicke parameter
\cite{LD}, which describes the coupling strength of ion 
number $i$ to motional-mode number $p$. 

The two-qubit gate infidelity due to inexact 
phase space closure at zero temperature 
is given by \cite{GREEN} 
\begin{align}
f = \Big(\frac{4}{5}\Big) \sum_{p=1}^N 
\big(|\alpha_p^i(\tau)|^2+|\alpha_p^j(\tau)|^2 \big),
\end{align}
where $i$, $j$ are qubit indices, 
$p$ is the motional-mode index, 
and $\alpha_p^i(\tau)$ is defined in (\ref{prior2}).
Inserting (\ref{prior1}) in (\ref{prior2}) with 
$k=0$, we obtain
\begin{align}
    \alpha_p^i(\tau) &= -\eta_p^i \sum_{n=1}^{N_A} A_n \int_0^{\tau} \text{sin}\Big(\frac{2\pi n}{\tau}t\Big) e^{i\omega_p t} dt \nonumber \\
    &= -\eta_p^i \sum_{n=1}^{N_A} A_n C_{np},
\end{align} 
where
\begin{align}
    C_{np} = \int_0^{\tau} \text{sin}\Big(\frac{2\pi n}{\tau}t\Big) e^{i\omega_p t} dt.
\end{align}
With this, $f$ now becomes
\begin{align}
    f = \Big(\frac{4}{5}\Big) \sum_{p=1}^N \Bigg\{\Big[(\eta_p^i)^2+(\eta_p^j)^2\Big] \Big( \sum_{n=1}^{N_A} A_n C_{np}\Big)^2 \Bigg\},
\end{align}
which can be rewritten as 
\begin{align}
    f=\Big(\frac{4}{5}\Big) \sum_{n=1}^{N_A} 
    \sum_{m=1}^{N_A} A_n F_{nm} A_m,
\end{align}
where
\begin{align}
    F_{nm} = \sum_{p=1}^N \Big[(\eta_p^i)^2+(\eta_p^j)^2\Big] C_{np}C_{mp},
\end{align}
and $m,n=1,2,\ldots,N_A$. 
In matrix notation, we can express the 
matrix $F$ according to 
\begin{align}
    F = CDC^T,
\end{align}
where
\begin{align}
    D_{pp'} = \Big[(\eta_p^i)^2+(\eta_p^j)^2\Big] \delta_{pp'}
\end{align}
is a positive, diagonal matrix. 
This shows that $F$ is positive semidefinite, 
i.e., it has only positive or zero eigenvalues. 
This is natural since $f$ is semidefinite as well.
The eigenvalues $\varphi_1 \leq \varphi_2 \leq 
\ldots \leq \varphi_{N_A}$ of $F$ determine the range of possible $f$ values. 
We denote the normalized eigenvector of $F$ corresponding to the eigenvalue $\varphi_l$ by $\hat{V}^{(l)}$. 
Since $F$ is real and symmetric, we have 
\begin{align}
    \label{f-subspace1}
    \Big[\hat{V}^{(l)} \Big]^T \cdot \hat{V}^{(l')} 
    = \delta_{l,l'} ,  
\end{align}
if $\varphi_l \neq \varphi_{l'}$, and in case 
of degeneracy, orthonormality can be achieved 
by orthogonalizing within the degenerate 
subspace. 
In order to obtain a relationship between infidelity and power, we compute the power in the space spanned by the eigenvectors of $F$, ranging from $\hat{V}^{(1)}$ to 
$\hat{V}^{(L_{\rm cut})}$, where $L_{\rm cut} \leq N_A$. 
Thus, for each choice of $L_{\rm cut}$, we obtain an (infidelity, power) pair. Infidelity and power in the reduced space are computed in the following way. \\
Expanding the vector $\vec{A}$ into the chosen subspace of the eigenvectors $\hat{V}^{(1)},\ldots,\hat{V}^{(L_{\rm cut})}$,
\begin{align}
\label{f-subspace2}  
    \vec{A} = \Omega_0\sum_{l=1}^{L_{cut}} \hat{B}_l \hat{V}^{(l)},
\end{align}
where $\Omega_0$ is the pulse amplitude and 
\begin{align}
    \label{f-norm}  
    \sum_{l=1}^{L_{\rm cut}} \hat{B}^2_l = 1, 
\end{align}
we obtain for the infidelity in the reduced space: 
\begin{align}
    f &= \Big(\frac{4}{5}\Big) \vec{A}^T F \vec{A} \nonumber\\ 
    &= \Big(\frac{4\Omega_0^2}{5}\Big) \Big[ \sum_{l=1}^{L_{\rm cut}} 
    \hat{B}_l \hat{V}^{(l)T}\Big] F \Big[ \sum_{l'=1}^{L_{\rm cut}} \hat{B}_{l'} \hat{V}^{(l')}\Big] \nonumber \\
    &= \Big(\frac{4\Omega_0^2}{5}\Big) 
    \sum_{l=1}^{L_{\rm cut}} \hat{B}_l^2 \varphi_l.
\label{infi-red-sp}
\end{align}
It remains to compute $\hat{B}$. 
We start from the expression for the entangling angle $\chi(\tau)$
\begin{align}
\chi(\tau) &= \vec{A}^T \cal K \vec{A} \nonumber \\
&= \Omega_0^2 \Big[ \sum_{l=1}^{L_{\rm cut}} 
\hat{B}_l \hat{V}^{(l)T}\Big] {\cal K}
\Big[ \sum_{l'=1}^{L_{\rm cut}} 
\hat{B}_{l'} \hat{V}^{(l')}\Big] \nonumber \\
&= \Omega_0^2\sum_{l,l'=1}^{L_{\rm cut}} \hat{B}_l S_{l,l'} \hat{B}_{l'},
\label{chi-eqn} 
\end{align}
where 
\begin{align}
    S_{l,l'} = \hat{V}^{(l)T} {\cal K} \hat{V}^{(l')},
\end{align}
and 
\begin{align}
    {\cal K}_{nm} &= \sum_{p=1}^{N}\eta^i_p\eta^j_p \int_0^{\tau} dt_2 \int_0^{t_2} dt_1  \text{sin}\Big(\frac{2\pi m}{\tau}t_1\Big) \nonumber \\
    &\text{sin}\Big(\frac{2\pi n}{\tau}t_2\Big) \text{sin}\Big[\omega_p (t_2-t_1)\Big].
\label{eq-kernel} 
\end{align}
Diagonalizing the  
reduced kernel matrix $S_{l,l'}$ in 
the subspace $1\leq l,l'\leq L_{\rm cut}$, 
we obtain its eigenvalues 
$\lambda_1,\lambda_2,\ldots, \lambda_{L_{\rm cut}}$.  
Finally, we find $\hat{B} = \vec{\nu}$, 
where $\vec{\nu}$ is the normalized eigenvector 
of $S$ corresponding
to the eigenvalue $\lambda_{\text{max}}$ 
with the largest absolute value.  
Thus, with (\ref{chi-eqn}), 
we accomplish the maximally entangling 
gate, $\chi(\tau) = \frac{\pi}{8}$, 
with the minimal $\Omega_0$ according to
\begin{align}
    \Omega_{0}= 
    \left( \frac{\pi}{8 |\lambda_{\text{max}}|}\right)^{1/2}.
    \label{omega0}
\end{align}
With (\ref{f-norm}), we find that the 
infidelity (\ref{infi-red-sp}) satisfies 
\begin{equation}
    \frac{4\Omega_0^2\varphi_1}{5} 
    \leq f \leq 
    \frac{4\Omega_0^2\varphi_{L_{\rm cut}}}{5} . 
\end{equation}
Therefore, 
\begin{equation}
    f_{\max} = \left(\frac{4\Omega_0^2}{5}\right)
    \varphi_{L_{\rm cut}}
\end{equation}
is a rigorous upper bound of the infidelity $f$. 

\section{Extended null-space protocol}
\label{sec:Sliding}

The extended null-space method presented here is an alternative way to generate efficient AMFM control pulses that satisfy $\alpha_p^i(\tau)\approx 0$ for all $i$ and $p$. Conceptually, the pulse construction relies on a procedure similar to that in (\ref{f-subspace1})-(\ref{f-norm}), and offers similar savings in the peak power levels of resulting pulses. Specifically, whereas $L_{\rm cut}$ controls the  number of eigenvectors of matrix $F$ included in the chosen solution subspace [see Supplementary Material 
Section~\ref{FinexactAMFM}, equation~(\ref{f-subspace2})], 
here, the subspace is formed only by 
eigenvectors of the matrix $\Gamma$, 
defined in (\ref{Gamma}), 
whose eigenvalues, in absolute 
magnitude, are smaller than some threshold $Z$. Thus, by adjusting the value of $Z$, we allow for additional eigenvectors of $\Gamma$, 
other then null-space vectors with eigenvalue zero, to be included in the solution subspace.

In the exact AMFM approach, described in detail in Ref.~\cite{AMFM},
the aim is to find the null space of the matrix $M$, 
i.e., the set of $N_0=N_A-N(K+1)$ linearly independent, nontrivial  
vectors $\vec{A}^{(\alpha)}$, $\alpha=1,\ldots,N_0$, that 
satisfy
\begin{align}
\label{prior3}
& - \bigg(\frac{\partial^k}{\partial \omega_p^k}\bigg) \eta_p^i 
\sum_{n=1}^{N_A} A_n \int_0^{\tau} \text{sin}\Big(\frac{2\pi n}{\tau}t\Big) e^{i\omega_p t} dt & \nonumber \\
&= \sum_{n=1}^{N_A} M_{mn}A_n =  M\vec{A} = 0,
\end{align}
where the matrix $M$ contains the stabilized phase-space closure conditions defined under Eq.~(\ref{prior2}).
In order for (\ref{prior3}) to have non-trivial solutions, 
we require that $N_A>N(K+1)$. 
Thus, in general, the rank-deficient, 
rectangular matrix $M$ has $N(K+1)$ 
non-zero eigenvalues and $N_0 = N_A-N(K+1)$ 
zero eigenvalues. This 
suggests to 
multiply (\ref{prior3}) from the left with the transpose, 
$M^T$, of $M$, 
which turns (\ref{prior3}) into 
\begin{equation}
\Gamma \vec A = 0, 
\label{prior5} 
\end{equation}
where 
\begin{equation} 
\Gamma=M^T M
\label{Gamma}
\end{equation} 
is a symmetric matrix. 
Section~S17 of Supplemental Material of Ref.~\cite{AMFM} 
shows that the operation of 
multiplication with $M^T$ does not change the structure of 
the solution space. 

Where the exact AMFM approach allows only the 
$N_0$ null-space vectors of $\Gamma$ with 
$\lambda_j=0$, $j=1,\ldots,N_0$, 
and satisfies (\ref{prior5})
exactly, here, we include additional eigenvectors in an extended ``null''-space of dimension 
larger than $N_0$ that 
correspond to non-zero eigenvalues of 
$\Gamma$. 
Diagonalizing 
$\Gamma$, we obtain its eigenvalues 
$\lambda_1,\lambda_2,\ldots, \lambda_{N_A}$, 
ordered according to the size of their absolute 
values, such that 
\begin{align}
    0\leq|\lambda_1| \leq|\lambda_2|\leq\ldots\leq|\lambda_{N_A}|.
\end{align}
Since the magnitude of each non-zero eigenvalue is proportional to the magnitude of violation of (\ref{prior5}), we 
define a parameter $Z$ (e.g., $Z=10^{-4}$) and 
include only eigenvalues whose absolute magnitude is 
smaller than $Z$ in the extended ``null''-space. 
With this definition of the 
extended ``null''-space, the power optimization procedure follows (\ref{chi-eqn})-(\ref{omega0}) with $N_0'$ in place of $L_{\rm cut}$, where $N_0'$, 
the dimension of the extended ``null''-space, 
is given by 
\begin{equation}
    N_0' = N_A - N(K+1) + N_{\rm ENS},
\end{equation}
where $N_{\rm ENS}$ is the number of eigenvectors of 
$\Gamma$ whose eigenvalues are greater than 0 but less than $Z$ in their absolute values, 
$N_{\rm ENS} \in [0,N(K+1)]$, and ENS stands for 
``Extended Null Space''. 
By allowing additional eigenvectors of 
$\Gamma$ into the solution subspace, we effectively relax the last line of (\ref{prior3}) into 
\begin{align}
    M\vec{A} \approx 0,
\end{align}
which yields control pulses with lower power than does the exact AMFM protocol. 

\section{Stabilized Gate Demonstration}
\label{sec:StabPulse}

\begin{figure}[t]
    \centering
\includegraphics[width=\linewidth]{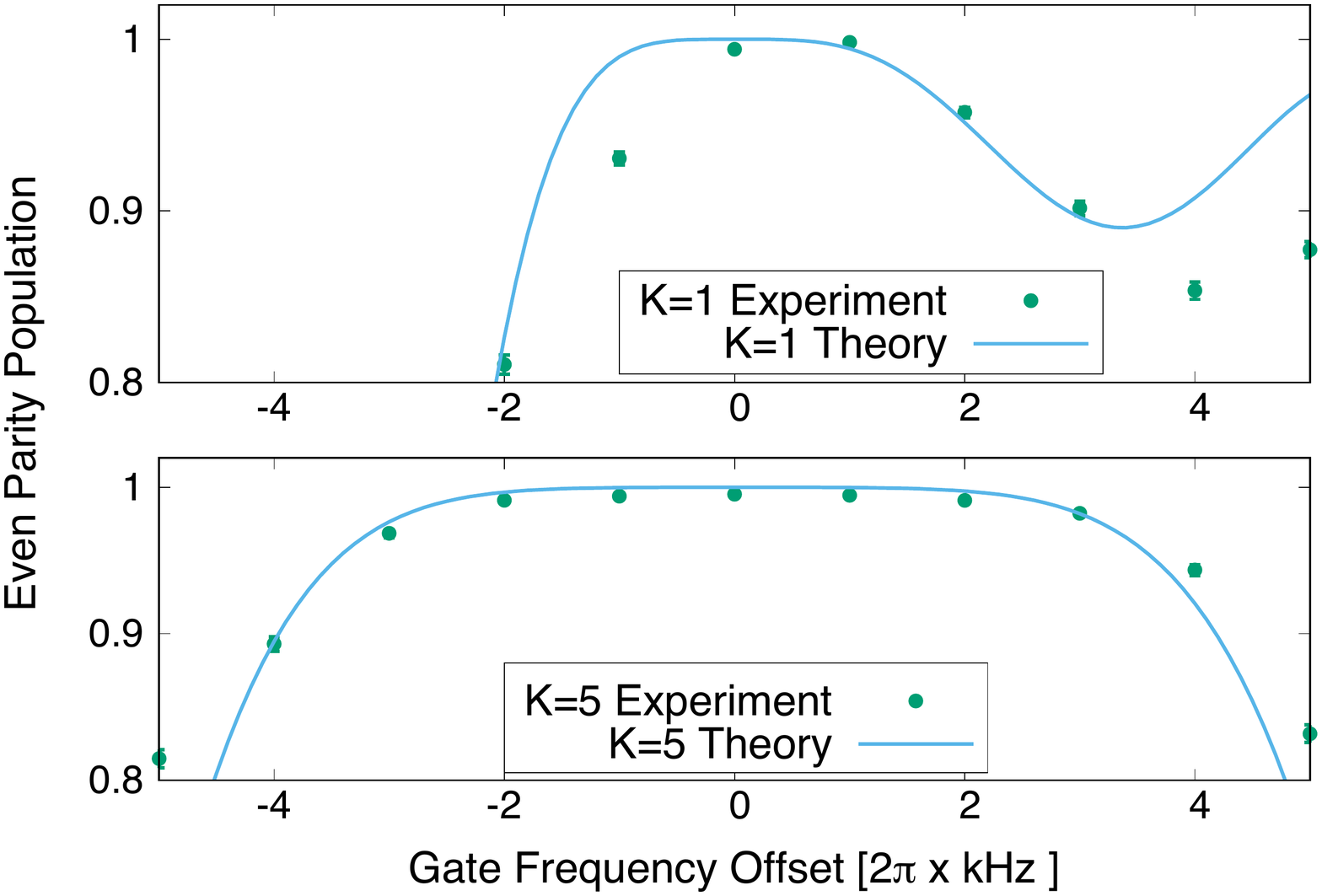}

\caption{\label{fig:stab} Experimental demonstration of the K-order stabilized two-qubit gates for qubit pair (4,5) on our seven-ion, five-qubit quantum computer. The ENS AMFM pulse sequence satisfying $f \leq 10^{-4}$ at no frequency offset with $K=1$ (top) and $K=5$ (bottom), corresponds to a maximally entangling gate. The experimentally measured even-parity population is plotted as a function of the gate frequency. The blue lines show the gate fidelity according to the analytical expression in (\ref{eq:infid}), valid in the low-error limit. The width of the detuning-robust region is greater for the the pulse with the higher stabilization order $K=5$ than for the pulse with $K=1$.}
\end{figure}

We demonstrate the benefit of stabilizing the gates with respect to fluctuations in the motional mode frequencies presented in the paper, by applying pulses with two different stabilization orders $K=1$ and $K=5$ to qubits (4,5) on our seven-ion, five-qubit quantum computer.  The two ENS AMFM pulses are designed for theoretical $f \leq 10^{-4}$ at zero gate frequency offset. To systematically control the detuning error, we offset the gate frequency from the original intended gate frequency, which is equivalent to uniformly offsetting the motional mode frequencies in the opposite direction. Figure \ref{fig:stab} shows the even-parity population $P_{\rm even} = P_{\rm 00} + P_{\rm 11}$, which is a proxy for the gate fidelity, as a function of gate frequency offset. We apply the pulses to the initial state $|00\rangle$ and measure the even-parity populations when $P_{\rm 00} \approx P_{\rm 11}$, akin to performing a maximally entangling gate. The experimentally measured values with their associated error bars are marked in green. The error bars are 1$\sigma$ confidence intervals, sampled from a binomial distribution, and each point represents 4000 experimental shots. The blue line shows the analytical fidelity
\begin{equation}
\bar F = 1-\frac{4}{5}\sum_p 
\left(|\alpha_{i,p}|^2 + |\alpha_{j,p}|^2 \right),
\label{eq:infid}
\end{equation}
which is valid in the low-infidelity limit \cite{GREEN}. The width of the detuning-robust region is greater for the the pulse with the higher stabilization order $K=5$ than for the pulse with $K=1$.

\section{Experimental Details}
\label{sec:ImpDet}

\begin{figure*}[h]
    \centering
    \includegraphics[width=0.85\linewidth]{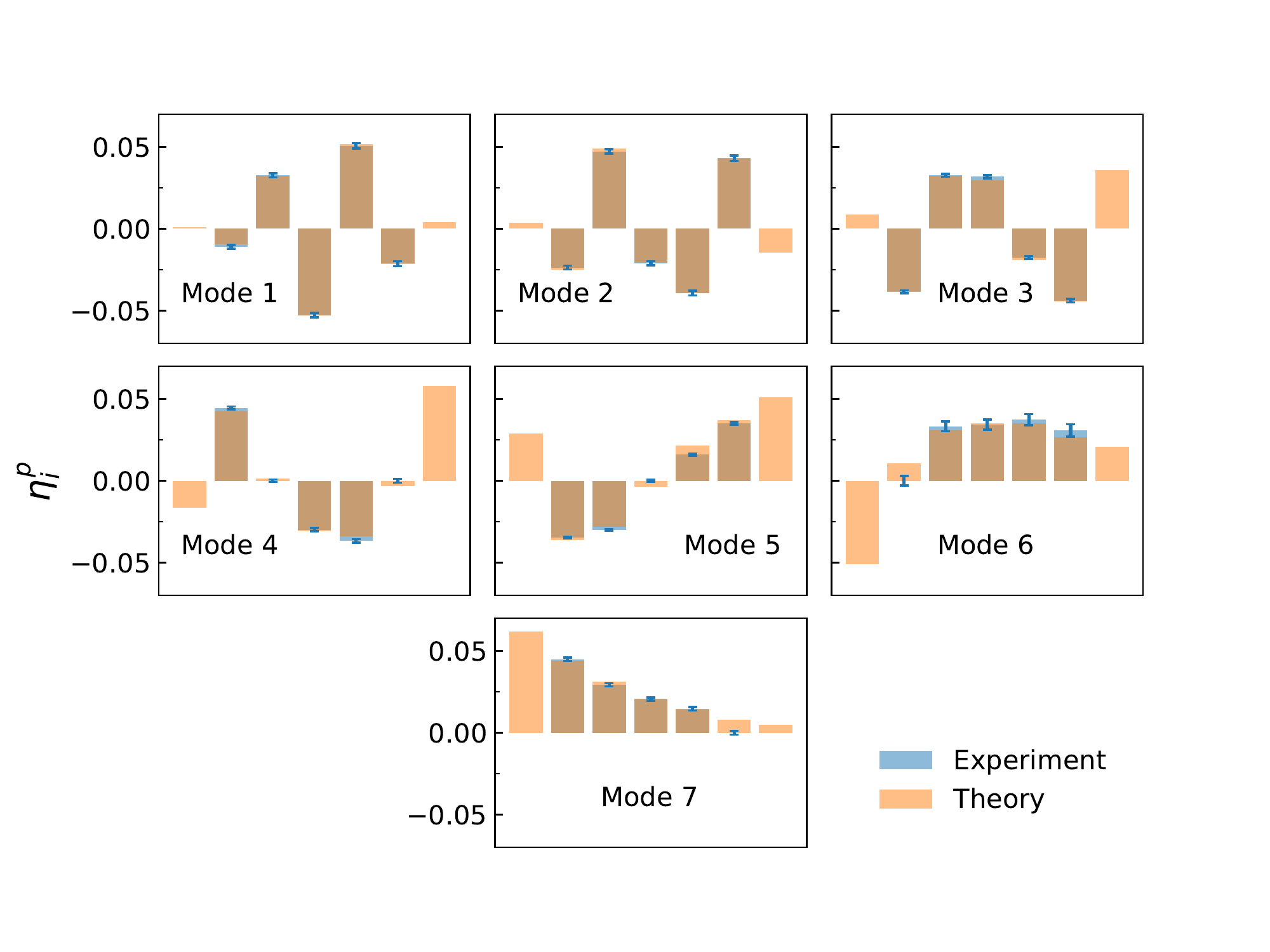}
    \caption{Comparison between experimentally
    extracted Lamb-Dicke parameters and theoretically
    fitted ones. The signs of the experimentally
    extracted Lamb-Dicke parameters are forced to be
    the same as the theoretically predicted ones since
    we can only extract the magnitude of the Lamb-Dicke
    parameters using sideband spectroscopy.
    The blue bars are experimental measurements. The blue error bars are statistical errors propagated through the fitting routines by bootstrap. The error bars do not account for systematic errors such as mode drift, heating and motional decoherence during the measurements.  
    Only the middle five ions are
    accessible by Raman lasers thus there is not
    any data for the two ions at the ends of the chain.
    The orange bars represent the fitted result
    using the simple theoretical model described in
    SM Sec.~\ref{sec:ImpDet}. The modes are indexed
    in the order of increasing mode frequency, as shown in Tb.~\ref{tab:ModeFs}.}
    \label{fig:eta}
\end{figure*}

In this section, we describe in detail the experimental setup and procedures. We encode the $\ket0$ and $\ket1$ qubit states in the two hyperfine levels $\ket{F=0,m_F=0}$, $\ket{F=1,m_F=0}$ of the ground state ${}^2S_{1/2}$ of ${}^{171}\rm Yb^+$. The qubit frequency is given by the hyperfine splitting $\omega_{HF}=2\pi \times 12.6428 \text{ GHz}$ and is magnetic field insensitive to first order \cite{Olmschenk07}. The ions are confined radially by an RF field at $\omega_{RF}=2\pi \times 23.83 \text{ MHz}$ and axially by a static field. The radial standard pseudo-frequency in the transverse gate direction, denoted X is $\omega_{7}=2\pi \times 3.054(1) \text{ MHz}$. In all of our experiments, we employ a two-step cooling process on the X modes with Doppler cooling to $\bar{n}\approx10$ and sideband cooling to $\bar{n}\approx0.1$. As described in the main text, the qubit transition can be driven coherently with a pair of counter-propagating Raman beams at 355 nm, controlled individually by a 32-channel AOM using RF pulses. We can adjust the beatnote of the Raman beams by changing the RF frequencies to address the qubit, the blue sideband or the red sideband transition. Finally, projective measurements are performed with state-dependent fluorescence on an array of photomultiplier tubes (PMT) with $98.4(1)\%$ readout fidelity for a two-qubit state and $95(1)\%$ readout fidelity for a five-qubit state.

To accurately determine the experimental Rabi frequencies for every two-qubit gate presented in Fig.~\ref{fig:exp}, we employ the following procedure. To eliminate the effect of beam alignments and laser power fluctuation on the Rabi frequencies, for each experiment, we measure the carrier Rabi frequencies of each qubit at the scale $s=1$. We then scan the scale to observe coherent population transfer between $\ket{00}$ and $\ket{11}$, a signature of the MS gate. From this scan, we can determine the scale for maximally entangling gate. We perform the gate at this scale with 10000 data points and correct for state-preparation and measurement (SPAM) errors in post-processing to obtain a good estimate of the scale, and therefore the Rabi rate $\Omega_0$.

A carefully designed protocol for accurately determining the motional mode frequencies as well as the Lamb-Dicke parameters $\eta_p^i$ is necessary to achieve a good agreement between predicted and observed Rabi rates needed for a maximally entangling gate. Here, we present the details of the protocol we implement in terms of experimental procedures and post processing steps.
Experimentally, we aim to measure the mode frequencies and the magnitude of the Lamb-Dicke parameters through sideband spectroscopy.

After ground-state cooling and initialization, we scan the frequencies across the blue sideband for each ion, one ion at a time, with square pulses at a fixed laser power and pulse duration before detecting the population in $|1\rangle$ state.
For each normal mode and each ion, we can define a sideband Rabi frequency
as $\Omega_i^p = \sqrt{\bar{n}+1}\Omega_i\eta_i^p$, where
$\Omega_i$ is the Rabi frequency of the qubit transition.
When $\Omega_i^p$ is much smaller than the splitting
between motional mode frequencies, the transitions with frequency
in the vicinity of the motional mode frequency $\omega_p$ can be modeled
as a two-level system and the final $|1\rangle$ state
population after a pulse with a duration $t_{\rm bsb}$ 
can be approximated as
\begin{equation}
    P_{|1\rangle} =
    \frac{{\Omega_i^p}^2}{{\Omega_i^p}^2+\delta\omega_p^2/4}
    \sin^2\left(t_{\rm bsb} \sqrt{{\Omega_i^p}^2+\frac{\delta\omega_p^2}{4}}\; \right) \;,
    \label{eq:p1}
\end{equation}
where $\delta\omega_p$ is the difference between the
mode frequency $\omega_p$ and the frequency of the sideband pulse.
Thus, with enough frequency points scanned near each
mode, we can fit the qubit $|1\rangle$ state population
as a function of frequency to (\ref{eq:p1}) and
extract $\omega_p$ and $|\Omega_i^p|$.
Note that we choose the pulse duration and the Rabi frequency
of the qubit transition with care so that $P_{|1\rangle} \lesssim 50\%$ to minimize the
effect of the thermal distribution of the mode phonon population.
Using the Rabi frequency of the qubit transition, which
can be measured in a straightforward fashion,
the magnitude of the Lamb-Dicke parameter is given
by $|\eta_i^p| = |\Omega_i^p|/\sqrt{\bar{n}+1}\Omega_i$.
After measuring them for all modes and addressable ions, we fit
the magnitudes and the mode frequencies to
a simple theoretical model of a linear ion chain and obtain theoretically
fitted $\eta_i^p$, including their signs. 
The model includes the Coulomb interaction between
ions and individual harmonic confinements for each ion
along the direction that is perpendicular to the chain.
The inter-ion spacing and the spring constants
for the harmonic confinements are the fit parameters.
An example of the fitted Lamb-Dicke parameters is
shown in Fig.~\ref{fig:eta}, which shows the
quality of the fit is excellent.
For increasingly small $\eta^p_i$ values, measuring them precisely and accurately becomes difficult, since the BSB Rabi frequencies $\Omega^p_i$ are smaller than the frequency resolution of the scan (which is $1.5$ kHz), and the measurements become dominated by RF noises of the trap.
Therefore, it is more
appropriate to use the theoretically fitted Lamb-Dicke
parameters for the pulse-shaping calculations.

\begin{figure*}[h]
    \centering
    \includegraphics[width=\linewidth]{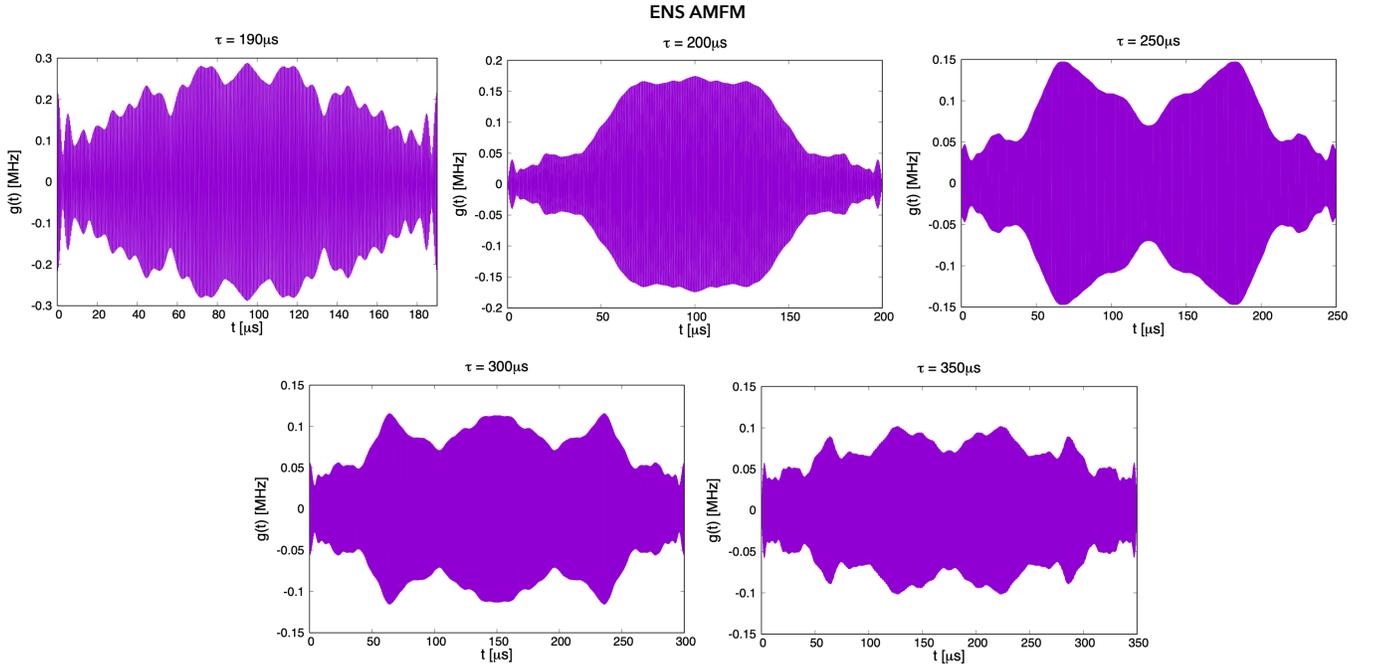}
    \caption{Pulse shapes for two-qubit gates on qubits 4, 5 computed according to the extended null-space AMFM protocol.}
    \label{ENSPulses}
\end{figure*}

The mode frequencies used to generate the pulses are reported in Table~\ref{tab:ModeFs}. Figures~\ref{ENSPulses} and \ref{ExactPulses} show the pulse shapes used to implement the two-qubit gates on our seven-ion, five-qubit trapped-ion quantum computer.
\begin{figure*}[h]
    \centering
    \includegraphics[width=\linewidth]{S6-exact.pdf}
    \caption{Pulse shapes for two-qubit gates on qubits 4, 5 computed according to the exact AMFM protocol.}
    \label{ExactPulses}
\end{figure*}

\begin{table}[h]
\begin{tabular}{cc}
 \hline
Mode &  $\omega_p/2\pi$ (MHz)\\ \hline \hline
  1 &    2.951 \\
  2 &    2.973 \\
  3 &    2.993 \\
  4 &    3.010 \\
  5 &    3.025 \\
  6 &    3.038 \\
  7 &    3.054 \\
    \hline
\end{tabular}
\caption{Mode frequencies of the motional modes of our seven-ion chain.}
\label{tab:ModeFs}
\end{table}
\end{document}